\documentclass{article}

\usepackage{hyperref}

\usepackage{amsmath}
\usepackage{amssymb}
\usepackage{seqsplit}
\usepackage{listings}
\usepackage{color}
\usepackage{algpseudocode}
\usepackage{algorithm}
\usepackage{array}
\usepackage{breakurl}
\usepackage{multirow}
\usepackage{rotating}
\usepackage{indentfirst}
\usepackage{mdwlist}
\usepackage{setspace}

\usepackage{booktabs}

\usepackage{graphicx}
\usepackage{color}
\usepackage{subcaption}

\usepackage[version=4]{mhchem}

\usepackage[T1]{fontenc}

\usepackage{mathtools}

\title{Most abundant isotope peaks and efficient selection on $Y=X_1+X_2+\cdots + X_m$}

\author{Patrick Kreitzberg\\
  University of Montana\\
  Department of Computer Science\\
  32 Campus Drive, Missoula, MT\\
  United States of America
  \and
  Kyle Lucke\\
  University of Montana\\
  Department of Computer Science\\
  32 Campus Drive, Missoula, MT\\
  United States of America
  \and
  Oliver Serang\\
  University of Montana\\
  Department of Computer Science\\
  32 Campus Drive, Missoula, MT\\
  United States of America\\
  oliver.serang@umontana.edu
}

\date{\today}

\begin{document}

\sloppy

\maketitle

\begin{abstract}
  \noindent The isotope masses and relative abundances for each element are
  fundamental chemical knowledge. Computing the isotope masses of a
  compound and their relative abundances is an important and difficult
  analytical chemistry problem. We demonstrate that this problem is
  equivalent to sorting $Y=X_1+X_2+\cdots+X_m$. We introduce a novel,
  practically efficient method for computing the top values in $Y$. 
  then demonstrate the applicability of this method by computing the
  most abundant isotope masses (and their abundances) from compounds
  of nontrivial size.
\end{abstract}

\section{Introduction}
Atoms of the same element may have a variable numbers of neutrons in
their nulcei. The number of neutrons in the nucleus of an atom
influences its mass. The relative abundances with which different
isotopes naturally occur is well established.\cite{lacki:isospec}

In compounds composed of several elements, finding the relative
abundances of the most prevalent isotopes of the compound (and the
respective masses at which these isotopes occur) is a difficult
combinatorial problem.

For small problems, this can be solved via brute force: it is possible
to compute all isotope masses and their respective abundances, sort
them in descending order of abundance, and then retrieving the isotope
peaks (we will refer to the mass and relative abundance of a
particular isotope as an isotope peak, because that is how they are
observed in mass spectrometry) with with greatest abundance; however,
the runtime of this brute-force approach is far too inefficient,
because it grows $\in \Omega(2^n)$.

\L\k{a}cki \emph{et al.}\cite{lacki:isospec} recently introduced a
statistical approach to this problem, by which the top isotope peaks
of a compound may be efficiently approximated. For each element, the
method generates data from the isotopologue (\emph{i.e.}, from the
contributions all possible isotopes from that element). For each
element, its possible contributions follow a multinomial distribution,
which \L\k{a}cki \emph{et al.} approximate using multivarite
Gau{\ss}ians and generate in descending order or probability. The
isotopologue contributions are combined over the relevant element
sets. At this point, they generate the isotope configurations of the
compound in descending order of probability, which is equivalent to
finding the $k$ most probable isotopologues of the molecule.

In this paper, we demonstrate that finding the top $k$ isotope peaks
of a compound composed of $m$ elements is equivalent to finding the
top $k$ values (and respective indices) in $Y=X_1+X_2+\cdots+X_m$,
where $X_i$ are vectors of length $n$ and $Y$ is the Cartesian product
of these vectors under the operator $+$. This problem, which is
important to other problems such as
max-convolution\cite{bussieck:fast} and max-product Bayesian
inference\cite{serang:fast, pfeuffer:bounded}, is the generalization
of the pairwise problem in which we compute the top $k$ values in
$C=A+B$.

Finding the top $k$ values in $C=A+B$ is nontrivial. In fact, there is
no known approach that sorts all $n^2$ values of $C$ faster than
naively computing and sorting them in
$O(n^2 \log(n^2))=O(n^2 \log(n))$\cite{bremner:necklaces}; however,
Frederickson \& Johnson demonstrated that the top $n$ values in $C$
can be computed $\in O(n \log(n))$\cite{frederickson:complexity}.
Frederickson \& Johnson generalized the problem to obtaining the
$k^{\text{th}}$ top value in a matrix which is sorted by both columns
and rows\cite{frederickson:generalized}.  A sorted matrix may be built
with $X_1 + X_2$, but the method presented assumes the matrix is
already of this form and does not take into account the work to produce
the matrix from the vectors.

This can be observed by sorting both $A$ and $B$ in descending order,
and then sparsely building a matrix of $A_i+B_j$. If $A'$ and $B'$
represent sorted vectors such that $A_1' \geq A_2' \geq \cdots$ and
$B_1' \geq B_2' \geq \cdots$, then the maximal value of $C$ is
$A_1'+B_1'$. The second largest value in $C$ is $\max(A_1'+B_2', A_2'+B_1')$.

W.l.o.g., we know that we will never insert $A_{i+1}'+B_j'$ into the
sorted result of top values in $C$ without first inserting the larger
(or equal) value $A_i'+B_j'$. Thus, each time a value from the matrix
$A_i'+B_j'$ is found to be the next largest in $C$, the subsequent
next largest value in $C$ may be in the unvisited existing neighbors
(row or column, but not diagonal) of previously unvisited
values. These considered values form a ``fringe'' around the indices
whose values are already inserted into the result. We find the next
value to insert into the result by finding the minimum value currently
in this fringe. Because $\leq 2$ values will be inserted into the
fringe in each iteration (\emph{i.e.}, if $A_i'+B_j$ was just inserted
into the result, then $A_{i+1}'+B_j'$ and $A_i'+B_{j+1}'$ will be
added into the fringe if they are in bounds). Thus, the minimum value
in the fringe can be updated sparsely by using a binary heap. Note
that only the indices and values comprising the fringe are the only
values stored, and that the full matrix, which would have space $n^2$
and thus runtime $\in \Omega(n^2)$, is never realized. The fringe can
never have size $\in \omega(n)$ (because in the worst-case scenario,
it moves $n$ steps up and $n$ steps to the right), and so each next
largest value in $C$ will have cost $\in O(\log(n))$. Computing the
top $k$ values in $C$ is thus $\in O(k \log(n))$.

In this paper, we first construct a direct generalization of
Frederickson \& Johnson's method to the problem of finding the top $k$
values in $Y=X_1+X_2+\cdots+X_m$. We then create a more efficient
method by generalizing Frederickson \& Johnson's $C=A+B$ method to the
case where $A$ and $B$ are arbitrary, heap data structures, and
compute the largest $k$ values in $Y$ by constructing a balanced
binary tree whose nodes each are one of these data structures.  This
method is then applied to finding the most abundant isotope peaks for
a given molecular formula.

\section{Methods}
\subsection{A direct $m$-dimensional generalization of Frederickson \& Johnson}
The direct generalization of Frederickson \& Johnson's method, which
closely resembles \L\k{a}cki \emph{et al.}'s approach to generating
the top isotope peaks\cite{lacki:isospec}, is straightforward: Instead
of a matrix, we have an $\mathbb{R}^m$ tensor.  As before, we store
only the current fringe, which is stored in a heap. In each iteration,
we remove the minimal value in the fringe and append that to the
result vector. Let this minimal value come from index $(i_1,i_2,\ldots
i_m)$. Now we insert the $m$ values from the neighbors of index
$(i_1,i_2,\ldots i_m)$ into the heap holding the fringe:
$(i_1+1,i_2,\ldots,i_m), (i_1,i_2+1,\ldots,i_m), \ldots
(i_1,i_2,\ldots,i_m+1)$. As with the two-dimensional method, it is
possible to store not only the $X_{1,i_1} + X_{2,i_2} + X_{3,i_3} +
\cdots X_{m,i_m}$ in the heap, but also store the index tuple from
which it came. This is shown in
Listing~\ref{alg:m-dim-generalization}.

This $m$-dimensional method will be substantially slower than the
two-dimensional method: the fringe in this version of the
$m$-dimensional method is a plane of width $\leq n$ and dimension
$m-1$, and thus can have size up to $O(n^{m-1})$.

 \lstset{language=Python,
   basicstyle=\ttfamily\footnotesize,
   keywordstyle=\color{blue}\ttfamily,
   stringstyle=\color{red}\ttfamily,
   commentstyle=\color{magenta}\ttfamily,
   morecomment=[l][\color{magenta}]{\#},
   breaklines=true,
   label=alg:maxindexheap
 }
\begin{lstlisting}[caption={{\tt MaxIndexHeap.py}: A max heap Python class.}]
import heapq

class MaxIndexHeap:
  def __init__(self, values_and_indices=[]):
    self.min_heap = [ (-a,b) for a,b in values_and_indices ]
    heapq.heapify(self.min_heap)
    self.descending_values_and_indices = []

  def insert(self, val_and_index):
    # Note: does not guard against double insertion of an index
    val, index = val_and_index
    heapq.heappush(self.min_heap, (-val, index))

  def pop_max_and_index(self):
    neg_val, index = heapq.heappop(self.min_heap)
    self.descending_values_and_indices.append( (-neg_val, index))
    return (-neg_val, index)

  def get_value_and_index_at_rank(self, rank):
    assert(rank >= 0 and rank < len(self.descending_values_and_indices))
    return self.descending_values_and_indices[rank]

  def __len__(self):
    return len(self.min_heap)
\end{lstlisting}

 \lstset{language=Python,
   basicstyle=\ttfamily\footnotesize,
   keywordstyle=\color{blue}\ttfamily,
   stringstyle=\color{red}\ttfamily,
   commentstyle=\color{magenta}\ttfamily,
   morecomment=[l][\color{magenta}]{\#},
   breaklines=true,
   label=alg:m-dim-generalization
 }
\begin{lstlisting}[caption={{\tt TensorCartesianSumHeap.py}: A generalization of Frederickson \& Johnson's method, which computes the top $k$ values of $Y=X_1+X_2+\cdots+X_m$.  This method uses instances of the MaxIndexHeap class (Listing~\ref{alg:maxindexheap}).}]
from MaxIndexHeap import *

class TensorCartesianSumHeap:
  def __init__(self, vectors):
    self.m = len(vectors)
    self.descending_vectors_and_unsorted_indices = [ sorted([(b,a) for a,b in enumerate(v)], reverse=True) for v in vectors ]
    max_value = sum([a for a,b in [v[0] for v in self.descending_vectors_and_unsorted_indices]])
    zero_tup = (0,)*self.m
    self.fringe = MaxIndexHeap([(max_value,zero_tup)])
    self.sorted_indices_in_fringe = set( [zero_tup] )

  def pop_max_and_index(self):
    val,index = self.fringe.pop_max_and_index()
    self.sorted_indices_in_fringe.remove(index)

    # insert neighbors from fringe if not already visited:
    for i in range(self.m):
      new_index = list(index)
      new_index[i] += 1
      new_index = tuple(new_index)

      desc_vec_i = self.descending_vectors_and_unsorted_indices[i]

      if new_index[i] < len(desc_vec_i) and new_index not in self.sorted_indices_in_fringe:
        old_val_at_index = desc_vec_i[new_index[i]-1][0]
        new_val_at_index = desc_vec_i[new_index[i]][0]
        new_val = val - old_val_at_index + new_val_at_index
        self.fringe.insert( (new_val, new_index) )
        self.sorted_indices_in_fringe.add( new_index )

    unsorted_index = tuple([ self.descending_vectors_and_unsorted_indices[i][j][1] for i,j in enumerate(index) ])
    return (val, unsorted_index)

\end{lstlisting}

\subsection{A hierarchical $m$-dimensional method for finding the top $k$ values of $Y$}
First, observe that in the two-dimensional case $C=A+B$, it is not
necessary for $A$ and $B$ to be sorted vectors; instead, it is
sufficient that $A$ and $B$ simply be max-heap data structures, from
which we can repeatedly request the next largest value. The method
thus runs similarly to the two-dimensional case: the top-left corner
of $C$ is computed via the maximal value of $A$ and the maximal value
of $B$. This inserts two values into the fringe: either the sum of the
largest value in $A$ and the second-largest value in $B$ or the sum of
the second-largest value in $A$ and the largest value in $B$. Neither
the full, sorted contents of $A$ nor the full, sorted contents of $B$
are needed.

We thus construct a balanced binary tree of these heap-like
structures:
\begin{align*}
  Y=&(X_1+X_2+\cdots+X_{\frac{n}{2}}) + (X_{\frac{n}{2}+1}+X_{\frac{n}{2}+2}+\cdots+X_n)\\
  =&(X_1+X_2+\cdots+X_{\frac{n}{4}}) + (X_{\frac{n}{4}+1}+X_{\frac{n}{4}+2}+\cdots+X_{\frac{n}{2}}) +\\
  &(X_{\frac{n}{2}+1}+X_{\frac{n}{2}+2}+\cdots+X_{\frac{3n}{4}}) + (X_{\frac{3n}{4}+1}+X_{\frac{3n}{4}+2}+\cdots+X_n\\
  =&\cdots\\
\end{align*}

\begin{figure}
\centering
 \includegraphics[width=4in]{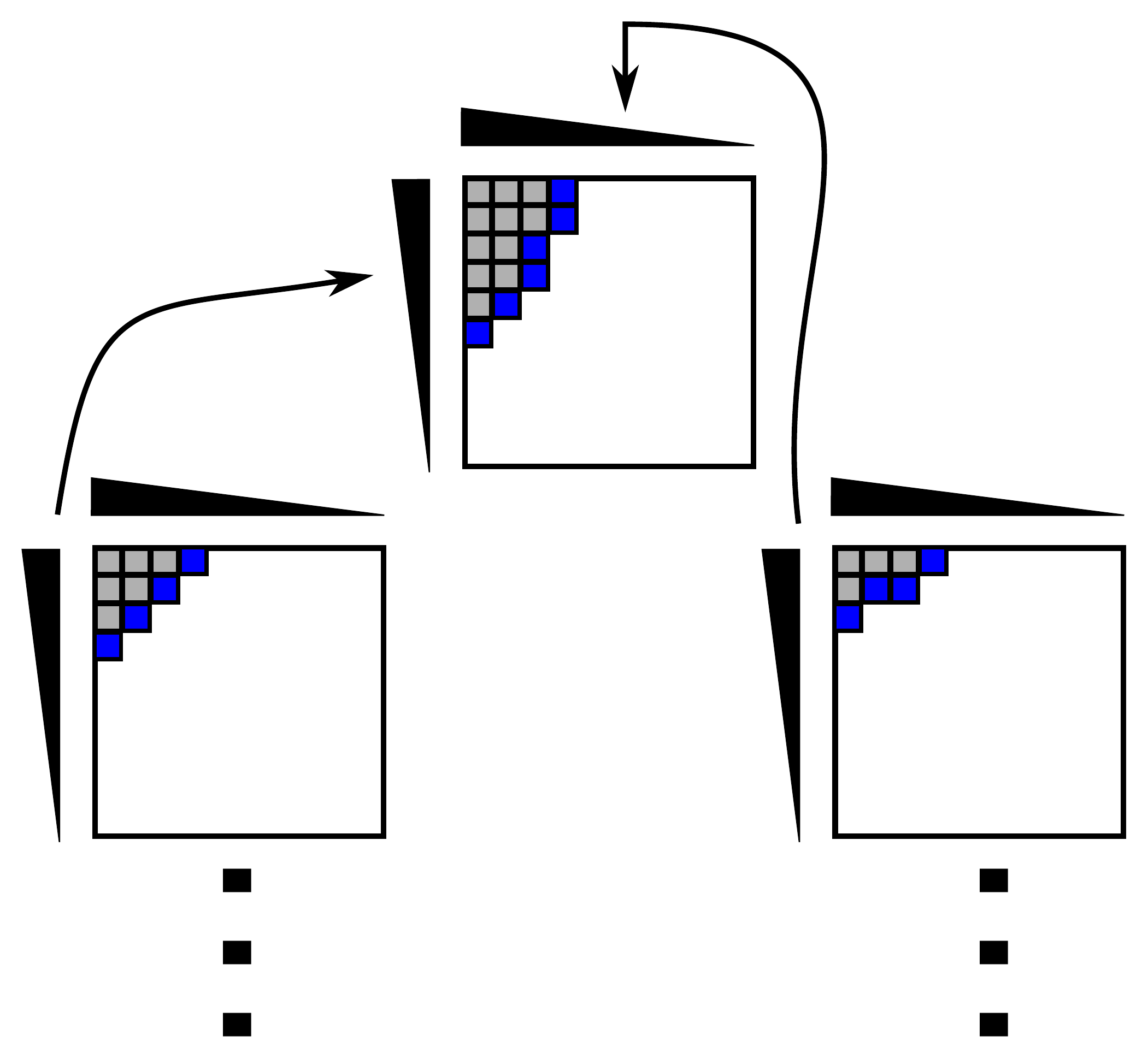}
 \caption{{\bf Illustration of the hierarchical method.} Pairwise
   Cartesian sum heaps (Listing~\ref{alg:cartesiansumpairheap}) are
   assembled into a balanced binary tree. The gray squares in each 2D
   grid represent values which have already been popped from that {\tt
     CartesianSumPairHeap} at the request of a parent node in the
   tree. When a value is popped from a child to the parent, it
   advances the corresponding margin along one axis of the parent's
   grid. The blue squares are values on the fringe, but which have not
   yet been popped. The child on the left has popped six values and
   currently has four values in its fringe; the row axis of the parent
   has six values that have been realized thus far. The child on the
   right has popped four values and currently has four values in its
   fringe; the column axis of the parent has four values that have
   been realized thus far. The indices from which the child popped are
   also included, enabling lookup of the index $(i_1,i_2,\ldots i_m)$
   was the next largest value
   $Y=X_{1,i_1}+X_{2,i_2}+\cdots+X_{m,i_m}$.
  \label{fig:tree-method}}
\end{figure}

Each heap-like structure is of the form $C=A+B$, where $A$ and $B$ are
heap-like structures (Figure~\ref{fig:tree-method}). The base case (at
the leaves) is achieved by simply using a binary heap of an input
vector (Listing~\ref{alg:hierarchical}).

 \lstset{language=Python,
   basicstyle=\ttfamily\footnotesize,
   keywordstyle=\color{blue}\ttfamily,
   stringstyle=\color{red}\ttfamily,
   commentstyle=\color{magenta}\ttfamily,
   morecomment=[l][\color{magenta}]{\#},
   breaklines=true,
   label=alg:cartesiansumpairheap
 }
\begin{lstlisting}[caption={{\tt CartesianSumPairHeap.py}: The CartesianSumPairHeap class is a heap-like structure which gets the next largest value either by pulling from one of its heap-like structure children, or from two sorted vectors contained in the class.  This class is used in the hierarchical method.}]
from MaxIndexHeap import *

def int_or_tuple_to_tuple(int_or_tuple):
  if type(int_or_tuple) is not tuple:
    return (int_or_tuple,)
  return int_or_tuple

class CartesianSumPairHeap:
  def __init__(self, heap_like_a, heap_like_b):
    self.heap_like_a = heap_like_a
    self.heap_like_b = heap_like_b

    self.number_remaining_elements = len(heap_like_a) * len(heap_like_b)

    max_a, index_a = heap_like_a.pop_max_and_index()
    max_b, index_b = heap_like_b.pop_max_and_index()
    index_a = int_or_tuple_to_tuple(index_a)
    index_b = int_or_tuple_to_tuple(index_b)
    max_value = max_a + max_b
    zero_tup = (0,0)

    self.fringe = MaxIndexHeap([(max_value, (zero_tup, index_a+index_b))])
    self.sorted_indices_in_fringe = set( [zero_tup] )

    self.descending_values_and_indices = []
    
  def insert_into_fringe(self, a_int_ind, b_int_ind):
    # check that indices are in bounds:
    if a_int_ind != len(self.heap_like_a.descending_values_and_indices) and b_int_ind != len(self.heap_like_b.descending_values_and_indices):
      matrix_index = (a_int_ind, b_int_ind)
      if matrix_index not in self.sorted_indices_in_fringe:
        new_val_a, new_index_a = self.heap_like_a.get_value_and_index_at_rank(a_int_ind)
        new_val_b, new_index_b = self.heap_like_b.get_value_and_index_at_rank(b_int_ind)
        new_index_a = int_or_tuple_to_tuple(new_index_a)
        new_index_b = int_or_tuple_to_tuple(new_index_b)
        self.fringe.insert( (new_val_a+new_val_b, (matrix_index, new_index_a+new_index_b)) )
        self.sorted_indices_in_fringe.add(matrix_index)
  
  def pop_max_and_index(self):
    self.number_remaining_elements -= 1

    val,(two_d_sorted_index,full_index) = self.fringe.pop_max_and_index()
    self.sorted_indices_in_fringe.remove(two_d_sorted_index)

    # insert neighbors of selected cell if not already visited:
    current_a = two_d_sorted_index[0]
    current_b = two_d_sorted_index[1]
    
    if len(self.heap_like_a)!=0 and current_a+1 == len(self.heap_like_a.descending_values_and_indices):
      self.heap_like_a.pop_max_and_index()
    if len(self.heap_like_b)!=0 and current_b+1 == len(self.heap_like_b.descending_values_and_indices):
      self.heap_like_b.pop_max_and_index()

    # at this point, right and down cells can be found in heap_like_a/b
    self.insert_into_fringe(current_a+1,current_b)
    self.insert_into_fringe(current_a,current_b+1)

    self.descending_values_and_indices.append( (val, full_index) )
    return (val, full_index)

  def get_value_and_index_at_rank(self, rank):
    assert(rank >= 0 and rank < len(self.descending_values_and_indices))
    return self.descending_values_and_indices[rank]

  def __len__(self):
    # hack because python only allows primitive int return value for __len__:
    if self.number_remaining_elements > 2**32-1:
      return 2**32-1
    return self.number_remaining_elements
\end{lstlisting}

 \lstset{language=Python,
   basicstyle=\ttfamily\footnotesize,
   keywordstyle=\color{blue}\ttfamily,
   stringstyle=\color{red}\ttfamily,
   commentstyle=\color{magenta}\ttfamily,
   morecomment=[l][\color{magenta}]{\#},
   breaklines=true,
   label=alg:hierarchical
 }
\begin{lstlisting}[caption={{\tt TreeCartesianSumHeap.py}: A hierarchical method for computing the top $k$ values of $Y=X_1+X_2+\cdots+X_m$ which uses objects of the class in Listing~\ref{alg:cartesiansumpairheap}.}]
from CartesianSumPairHeap import *

class TreeCartesianSumHeap:
  def __init__(self, vectors):
    vectors_and_indices = [ [(b,a) for a,b in enumerate(v)] for v in vectors ]

    # build a balanced binary tree and store the root:
    current_layer = [ MaxIndexHeap(v) for v in vectors_and_indices ]
    while len(current_layer) > 1:
      next_layer = []
      for i in range(len(current_layer) // 2):
        next_layer.append( CartesianSumPairHeap(current_layer[2*i], current_layer[2*i+1]) )
      if len(current_layer)%2 == 1:
        next_layer.append( current_layer[-1] )
      current_layer = next_layer
    self.root = current_layer[0]

  def pop_max_and_index(self):
    val,index = self.root.pop_max_and_index()
    return (val,int_or_tuple_to_tuple(index))

\end{lstlisting}

\subsection{Computing the most abundant isotope peaks}
Let carbon be represented as the vector $C=(\log(0.9892),
\log(0.0108)) = (-0.0108,-4.5282) $ and hydrogen as the vector
$H=(\log(0.9999),\log(0.0001)) = (-0.0001,
-9.2103)$.\cite{lacki:isospec} Propane, \ce{C3H8}, has abundances
composed of $Y=C+C+C+H+H+H+H+H+H+H+H$. To reduce the number of vectors
from 11 to two, a multinomial for each element is computed to find the
contributions from all isotopes. This multinomial is represented as a
vector which takes the place of all vectors for that element. Thus,
propane could be estimated as the sum of two multinomials, which are
encoded as vectors: one for $C$ and one for $H$. Molecules that
consist of any amounts of hydrogen, carbon, nitrogen, oxygen, sulfur,
\emph{etc.} can solved using vectors: $H, C, N, O, S\ldots$.  The most
abundant isotope peaks are found via the top values in $Y$.

The specific isotopes to which these abundances correspond (and from
which we can compute the masses that correspond to each abundance) can
be computed easily from the tuple indices. Python code implementing
the hierarchical method to calculate the most abundant isotope peaks
is shown in Listing~\ref{alg:abundant-isotope-calc}.

 \lstset{language=Python,
   basicstyle=\ttfamily\footnotesize,
   keywordstyle=\color{blue}\ttfamily,
   stringstyle=\color{red}\ttfamily,
   commentstyle=\color{magenta}\ttfamily,
   morecomment=[l][\color{magenta}]{\#},
   breaklines=true,
   label=alg:abundant-isotope-calc
 }
\begin{lstlisting}[caption={{\tt theodolite.py}: A method for computing the most abundant isotope peaks using the hierarchical $m$-dimensional method for finding the top $k$ values of $Y$.}]
import numpy
from collections import defaultdict
import sys
from scipy.special import gammaln
from functools import reduce

from TreeCartesianSumHeap import *
from TensorCartesianSumHeap import *
from BruteForceCartesianSumHeap import *

def read_in_elements(element_file):
  element_to_masses_and_abundances = {}
  for line in open(element_file):
    words = line.split()
    element_abbrev = words[0]
    tuples = [ tuple([float(x) for x in w.split(',')]) for w in words[1:] ]
    element_to_masses_and_abundances[element_abbrev] = tuples
  return element_to_masses_and_abundances

def parse_molecular_formula(formula):
  element_to_count = defaultdict(int)
  elements_and_counts = formula.split(',')
  for words in elements_and_counts:
    e_and_count = words.split(':')
    if len(e_and_count) == 1:
      e = e_and_count[0]
      count = 1
    else:
      e, count = e_and_count
      count = int(count)
    element_to_count[e] += count
  return element_to_count

def log_multinomial(log_probs, tup):
  result = gammaln(sum(tup)+1)
  for x in tup:
    result -= gammaln(x+1)
  for i in range(len(log_probs)):
    result += log_probs[i]*tup[i]
  return result

def tuples_with_fixed_sum(b, n):
    masks = numpy.identity(b, dtype=int)
    for c in itertools.combinations_with_replacement(masks, n): 
        yield tuple(sum(c))

# turn masses/log_abundances from single element into all combinations:
def element_to_multinomial_masses_and_log_abundances(element_masses, element_log_abundances, trials):
  n = len(element_masses)
  mass_results = []
  log_abundance_results = []
  for tup in tuples_with_fixed_sum(n, trials):
    mass = sum([ tup[i]*element_masses[i] for i in range(n) ])
    log_prob = log_multinomial(element_log_abundances, tup)
    mass_results.append(mass)
    log_abundance_results.append(log_prob)
  return mass_results, log_abundance_results

def usage():
  print( 'USAGE: <element file> <molecular formula e.g. H:2,S,O:4> <number_of_peaks> [tree(DEFAULT) OR tensor OR bruteforce]' )
  exit(1)

def main(args):
  if len(args) != 3:
    if len(args) != 4 or len(args) == 4 and args[3] not in ('tree', 'tensor', 'bruteforce'):
      usage()

  element_to_isotop_mass_and_abundance = read_in_elements(args[0])
  element_to_count = parse_molecular_formula(args[1])
  k = int(args[2])
  
  if len(args) == 4:
    mode = args[3]

  mass_vectors = []
  abundance_vectors = []
  for e,c in element_to_count.items():
    element_masses = [a for a,b in element_to_isotop_mass_and_abundance[e]]
    element_log_abundances = [numpy.log(b) for a,b in element_to_isotop_mass_and_abundance[e]]
    isotopologue_masses, isotopologue_log_abundances = element_to_multinomial_masses_and_log_abundances(element_masses, element_log_abundances, c)
    mass_vectors.append( isotopologue_masses )
    abundance_vectors.append( isotopologue_log_abundances )

  total_heap_size = reduce( (lambda x, y: x * y), [len(mv) for mv in mass_vectors])
  k = min(k, total_heap_size)

  if len(args) != 4 or mode == 'tree':
    cartesian_heap = TreeCartesianSumHeap(abundance_vectors)
  elif mode == 'tensor':
    cartesian_heap = TensorCartesianSumHeap(abundance_vectors)
  else:
    cartesian_heap = BruteForceCartesianSumHeap(abundance_vectors)
 
  log_abundance_and_index = [ cartesian_heap.pop_max_and_index() for i in range(k) ]
  print('Largest values in X_1*X_2*...:')
  print( '{:8} {:8} {:8}'.format('log(prob)', 'prob', 'mass') )
  for i in range(k):
    log_abund, ind = log_abundance_and_index[i]
    mass = sum([ mass_vectors[ell][j] for ell,j in enumerate(ind) ])
    print( '{:8f} {:8f} {:8f}'.format(log_abund, numpy.exp(log_abund), mass) )
  print()

  mass_to_total_abundance = defaultdict(float)
  for i in range(k):
    log_abund, ind = log_abundance_and_index[i]
    mass = sum([ mass_vectors[ell][j] for ell,j in enumerate(ind) ])
    mass_to_total_abundance[mass] += numpy.exp(log_abund)

  print('Most abundant peaks:')
  print( '{:8} {:8}'.format('prob', 'mass') )
  for a,m in sorted([ (b,a) for a,b in mass_to_total_abundance.items() ], reverse=True):
    print( '{:8f} {:8f}'.format(a,m) )

if __name__ == '__main__':
  main(sys.argv[1:])
\end{lstlisting}

\section{Results}
\subsection{Efficient isotope peak computation}
On a fake compound composed
of\\ \ce{Cl800V800He800C800H800N800O100S6Cu800Ga800Ag800Tl800Ne800}
with the cost of computing the multinomial isotopologues not included
(this cost was identical for both methods and required $\approx1$
second), computing the top 512 isotope beaks via a {\tt
  TensorCartesianSumHeap.py}-based implementation took 0.002984046
seconds, while the {\tt TreeCartesianSumHeap.py}-based implementation
took 0.00146198272 seconds.

\subsection{Time and memory use on arbitrary $Y=X_1+X_2+\cdots+X_m$}
\begin{figure}
\centering
\begin{tabular}{cc}
  \includegraphics[width=2.25in]{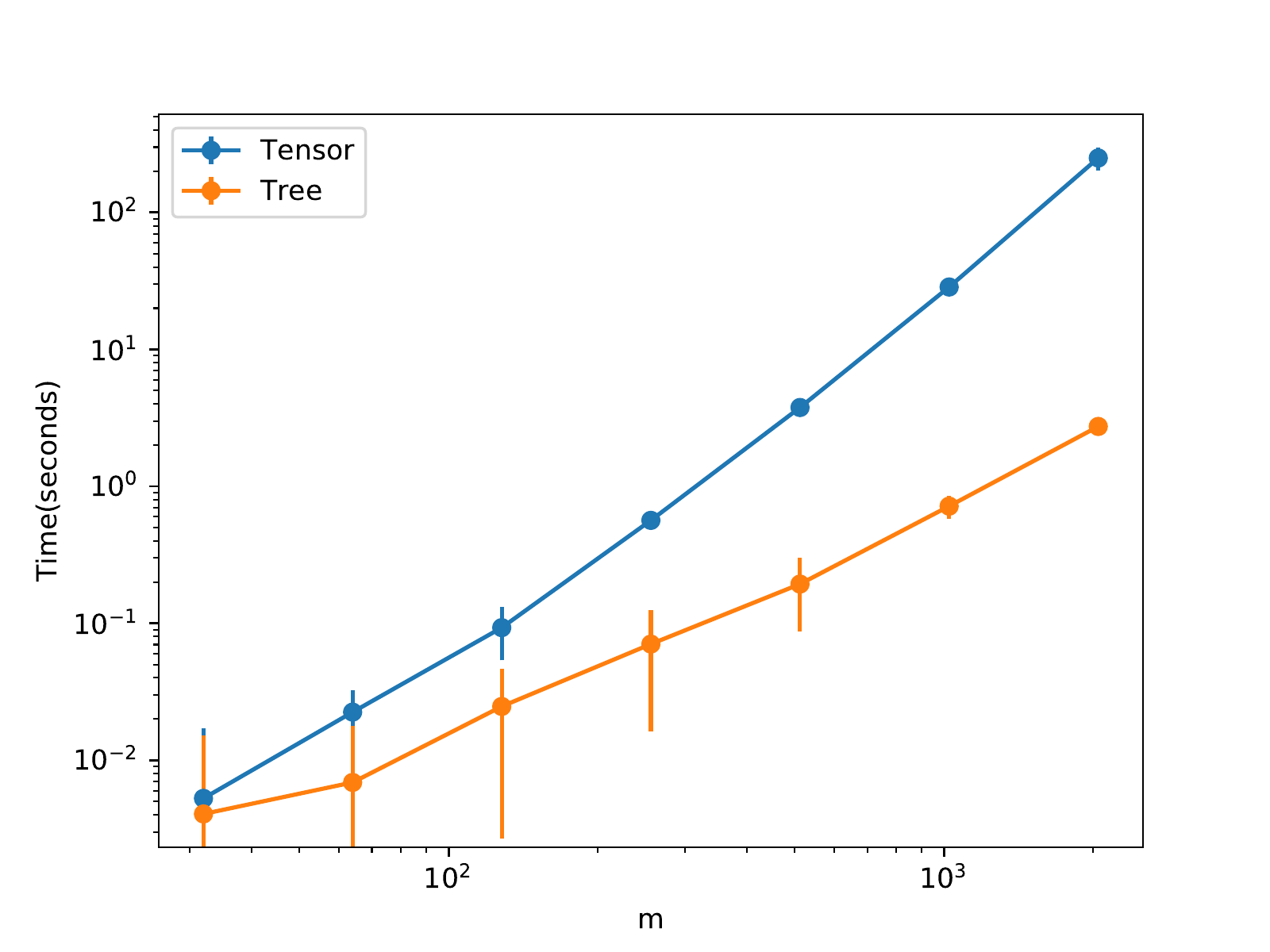} & \includegraphics[width=2.25in]{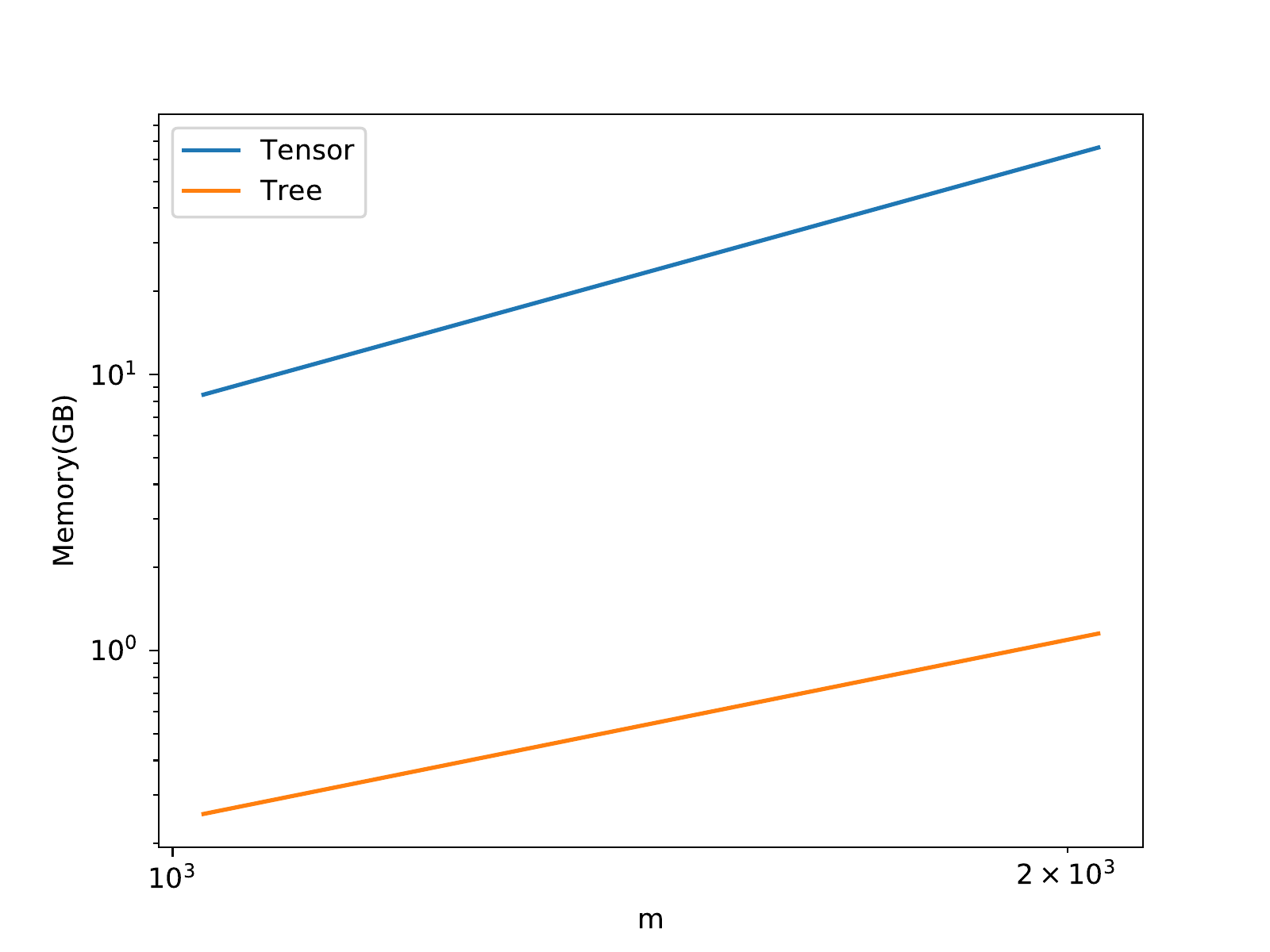}\\
\end{tabular}
\caption{{\bf Time and memory usage of {\tt TensorCartesianSumHeap.py}
    and {\tt TreeCartesianSumHeap.py}.} Problems of different size $m$
  and with $k=n=m$ are timed (left panel) and memory usage (right
  panel) are plotted. Memory usage of 0GB is due to a low enough
  footprint that the {\tt ps} command could not estimate it;
  therefore, we only show $m\in\{1024,2048\}$, for which both methods
  had nonzero memory usage. The growing gap in both of these log-log
  plots shows a nonlinear speedup and nonlinear memory benefit. At
  $m=2048$, the memory usage of {\tt TensorCartesianSumHeap.py} is
  66.3GB while the proposed hierarchical method, {\tt
    TreeCartesianSumHeap.py}, uses only 1.15GB.
  \label{fig:time-and-memory}}
\end{figure}

Problems of various sizes $m$ were run with vector length $n=m$,
retrieving the top $m$ values in $Y$. The brute-force approach was not
considered, for efficiency reasons. Time and memory use of {\tt
  TensorCartesianSumHeap.py} and {\tt TreeCartesianSumHeap.py} on a
dual Xeon with 128GB of RAM are shown in
Figure~\ref{fig:time-and-memory}.

\section{Discussion}
As $m$ increases, the method we introduce here is far more time
efficient, but more importantly, far more space efficient than direct
$m$-dimensional generalization of Frederickson \& Johnson.

Although the approach we propose here does have benefit to computation
of intense isotope peaks, the limited number of elements (currently at
$m=118$) benefits only slightly from our approach. Furthermore, it is
rare for many elements to be combined in a single compound. Our
demonstration implementation generates multinomials naively, unlike
\L\k{a}cki \emph{et al.}; the longer runtime from this unnecessary
na\"{i}vete in preprocessing mutes the speedup of the hiercharical
method for elements with several isotopes that are probable; however,
it is possible to use this hierarchical approach but with multinomials
generated non-naively as \L\k{a}cki \emph{et al.} do, which would
likely achieve a modest speedup over their current approach.

However, there are cases outside of computation of intense isotope
peaks, in which the hierarchical method we propose would yield large
practical benefit. These include \emph{maximum a posteriori} Bayesian
inference on dependencies of the form
$Y=X_1+X_2+\cdots+X_m$\cite{serang:probabilistic,pfeuffer:bounded}. Operations
research applications include financial markets, \emph{e.g.},
retrieving the $k$ lowest overall bids in each sector of supply lines
for a product.

Using a language like javascript, this method can be parallelized
easily by parallelizing heap pop operations in nodes at the same layer
of the tree. If enough hardware parallelism is available, the runtime
of a full propagation through the tree would thus be the height of the
tree, which is $\in \Theta(\log(m))$ pop operations from fringe
heaps. Each of these fringe heap pop operations is $\in O(\log(n))$,
and thus the runtime would be $\in O(\log(m)\log(n))$.

\section{Availability}
Python source code from this method is available at
\url{https://bitbucket.org/seranglab/theodolite/} (MIT license, free
for both academic and commercial use).

\end{document}